\documentclass[preprint,12pt,authoryear]{elsarticle}

\usepackage{hyperref}
\usepackage[switch]{lineno}
\usepackage[toc]{appendix}

\def\tsc#1{\csdef{#1}{\textsc{\lowercase{#1}}\xspace}}
\tsc{WGM}
\tsc{QE}

\usepackage{commath}
\usepackage{ragged2e}
\usepackage{multirow}
\usepackage{amsmath}
\usepackage{mathtools}
\usepackage{url}
\usepackage{balance}
\usepackage{booktabs}
\usepackage{listings}
\usepackage{algorithm}
\usepackage{algpseudocode}
\usepackage{caption}
\usepackage{cleveref}
\usepackage{xspace}

\crefformat{section}{\S#2#1#3}
\crefname{figure}{Figure}{Figures}
\crefname{appendix}{Appendix}{Appendices}
\crefname{table}{Table}{Tables}
\crefname{algorithm}{Algorithm}{Algorithms}
\crefname{listing}{Listing}{Listings}
\crefname{theorem}{Theorem}{Theorems}
\crefname{thm}{Theorem}{Theorems}
\crefname{lemma}{Lemma}{Lemmata}
\crefname{equation}{Eqt.}{Eqts.}
\crefformat{Grammar}{Grammar #1}

\algrenewcommand\alglinenumber[1]{\tiny #1:}
\usepackage[most]{tcolorbox}

\usepackage{soul}

\newcommand{\code}[1]{{\small\texttt{#1}}\xspace}
\newcommand{\errorcode}[1]{{\texttt{#1}}\xspace}

\newcommand{\domain}[1]{\textcolor{blue}{#1}}

\newtcbtheorem{Summary}{\bfseries Summary of RQ}{enhanced,
	coltitle=black,
	top=0.1in,
	attach boxed title to top left=
	{xshift=1.5em,yshift=-\tcboxedtitleheight/2},
	boxed title style={size=small,colback=lightgray}
}{summary}

\newtcbtheorem{Summary_data_charac}{Summary}{enhanced,
	coltitle=black,
	top=0.1in,
	attach boxed title to top left=
	{xshift=1.5em,yshift=-\tcboxedtitleheight/2},
	boxed title style={size=small,colback=gray}
}{Summary_data_charac}

\newtcolorbox[auto counter]{summary}[1][]{title={\bfseries Summary},enhanced,
	coltitle=black,
	top=0.17in,
	attach boxed title to top left=
	{xshift=1.5em,yshift=-\tcboxedtitleheight/2},
	boxed title style={size=small,colback=lightgray},#1}

\newtcolorbox[auto counter]{summary_PQ}[1][]{title={\bfseries Summary of Preliminary Study},enhanced,
	coltitle=black,
	top=0.17in,
	attach boxed title to top left=
	{xshift=1.5em,yshift=-\tcboxedtitleheight/2},
	boxed title style={size=small,colback=lightgray},#1}

\newtcolorbox[auto counter]{summary_RQ1}[1][]{title={\bfseries Summary of RQ1},enhanced,
	coltitle=black,
	top=0.17in,
	attach boxed title to top left=
	{xshift=1.5em,yshift=-\tcboxedtitleheight/2},
	boxed title style={size=small,colback=lightgray},#1}

\newtcolorbox[auto counter]{summary_RQ2}[1][]{title={\bfseries Summary of RQ2},enhanced,
	coltitle=black,
	top=0.17in,
	attach boxed title to top left=
	{xshift=1.5em,yshift=-\tcboxedtitleheight/2},
	boxed title style={size=small,colback=lightgray},#1}

\newtcolorbox[auto counter]{summary_RQ3}[1][]{title={\bfseries Summary of RQ3},enhanced,
	coltitle=black,
	top=0.17in,
	attach boxed title to top left=
	{xshift=1.5em,yshift=-\tcboxedtitleheight/2},
	boxed title style={size=small,colback=lightgray},#1}

\newtcolorbox[auto counter]{summary_RQ4}[1][]{title={\bfseries Summary of RQ4},enhanced,
	coltitle=black,
	top=0.17in,
	attach boxed title to top left=
	{xshift=1.5em,yshift=-\tcboxedtitleheight/2},
	boxed title style={size=small,colback=lightgray},#1}

\newtcolorbox[auto counter]{summary_RQ5}[1][]{title={\bfseries Summary of RQ5},enhanced,
	coltitle=black,
	top=0.17in,
	attach boxed title to top left=
	{xshift=1.5em,yshift=-\tcboxedtitleheight/2},
	boxed title style={size=small,colback=lightgray},#1}

\newcommand{\pqone}{\textit{PQ1: What is the popularity of Airflow-related questions on Stack Overflow?}}
\newcommand{\pqtwo}{\textit{PQ2: How difficult are Airflow-related questions?}}

\newcommand{\rqone}{\textit{What are the types of challenges developers face in using Airflow?}}
\newcommand{\rqtwo}{\textit{What are the root causes of the challenges?}}
\newcommand{\rqthree}{\textit{What are the types of online sources developers refer to in Airflow-related posts?}}

\newcommand{\rcone}{Incorrect Workflow Configuration}
\newcommand{\rctwo}{Complex Environmental Configuration}
\newcommand{\rcthree}{Insufficient Basic Knowledge of Airflow}
\newcommand{\rcfour}{Limitations of Airflow}
\newcommand{\rcfive}{Insufficient Knowledge of the External Systems}
\newcommand{\rcsix}{Incorrect or Sub-optimal Operator Choice}
\newcommand{\rcseven}{Missing or Incorrect Dependency}
\newcommand{\rceight}{Lack of knowledge in Programming and Devops}
\newcommand{\rcnine}{Complex Workflow}
\newcommand{\rcten}{Breaking Changes in Airflow}

\usepackage{tikz}
\usetikzlibrary{shapes.geometric,calc}

\newcommand\score[2]{
	\pgfmathsetmacro\pgfxa{#1+1}
	\tikzstyle{scorestars}=[star, star points=5, star point ratio=2.25, draw,inner sep=0.15em,anchor=outer point 3]
	\begin{tikzpicture}[baseline]
	\foreach \i in {1,...,#2} {
		\pgfmathparse{(\i<=#1?"yellow":"gray")}
		\edef\starcolor{\pgfmathresult}
		\draw (\i*1em,0) node[name=star\i,scorestars,fill=\starcolor]  {};
	}
	\pgfmathparse{(#1>int(#1)?int(#1+1):0}
	\let\partstar=\pgfmathresult
	\ifnum\partstar>0
	\pgfmathsetmacro\starpart{#1-(int(#1))}
	\path [clip] ($(star\partstar.outer point 3)!(star\partstar.outer point 2)!(star\partstar.outer point 4)$) rectangle 
	($(star\partstar.outer point 2 |- star\partstar.outer point 1)!\starpart!(star\partstar.outer point 1 -| star\partstar.outer point 5)$);
	\fill (\partstar*1em,0) node[scorestars,fill=yellow]  {};
	\fi
	
	,\end{tikzpicture}
}

{ \newcommand{\mynote}[2]{\textcolor{red}{
			\fbox{\bfseries\sffamily\scriptsize#1}
			{\small$\blacktriangleright$\textsf{\emph{#2}}$\blacktriangleleft$}}}}
{ \newcommand{\mynote}[2]{}}

\usepackage{subfigure}
\usepackage{footnote}
\usepackage{enumerate}
\usepackage{lipsum}
\usepackage{afterpage}
\usepackage{pdflscape}
\usepackage{lscape}
\usepackage{rotating}
\usepackage{booktabs, tabularx}

\usepackage{caption}
\usepackage{appendix}

\newcommand{\TotalQuestionPostsScoreGteZero}{{9,591}\xspace}

\newcommand{\ChallengeEnvSetup}{{18.3\%}\xspace}

\newcommand{\ChallengeBasicSetup}{{4.8\%}\xspace}
\newcommand{\ChallengeDeployment}{{4.1\%}\xspace}
\newcommand{\ChallengeModuleManagement}{{3\%}\xspace}

\newcommand{\ChallengeWorkflowDefinition}{{36.6\%}\xspace}

\newcommand{\ChallengeTaskConfiguration}{{20.1\%}\xspace}
\newcommand{\ChallengeIntegrationWithTheExternalSystems}{{9.6\%}\xspace}
\newcommand{\ChallengeSharingDataBetweenTasks}{{4.8\%}\xspace}
\newcommand{\ChallengeDesignoftheWorkflow}{{2.1\%}\xspace}

\newcommand{\ChallengeWorkflowExecution}{{23.8\%}\xspace}

\newcommand{\ChallengeScheduling}{{13.6\%}\xspace}
\newcommand{\ChallengeTaskDependency}{{10.2\%}\xspace}

\newcommand{\ChallengeWorkflowManagement}{{15.4\%}\xspace}
\newcommand{\ChallengeLoggingandMonitor}{{8.5\%}\xspace}
\newcommand{\ChallengeWorkflowMaintenance}{{4\%}\xspace}
\newcommand{\ChallengeTesting}{{1.4\%}\xspace}
\newcommand{\ChallengeMigration}{{1.5\%}\xspace}

\newcommand{\ChallengeSecurity}{{2.6\%}\xspace}
\newcommand{\ChallengeAdoption}{{1.2\%}\xspace}
\newcommand{\ChallengeOptimization}{{2.1\%}\xspace}
\newcommand{\URLReferences}{{16,059}\xspace}

\newcommand{\Airflow}{{Airflow}\xspace}

\begin{document}
\let\WriteBookmarks\relax
\def\floatpagepagefraction{1}
\def\textpagefraction{.001}






\title{An Empirical Study of Developers' Challenges in Implementing Workflows as Code:\\ A Case Study on Apache Airflow}                      

\author[inst1]{Jerin Yasmin}
\author[inst1]{Jiale Wang}
\author[inst1]{Yuan Tian}
\author[inst1]{Bram Adams}

\affiliation[inst1]{organization={School of Computing, Queen's University},
            city={Kingston},
            state={ON},
            country={Canada}}






\begin{abstract}
The Workflows as Code paradigm is becoming increasingly essential to streamline the design and management of complex processes within data-intensive software systems. These systems require robust capabilities to process, analyze, and extract insights from large datasets. Workflow orchestration platforms such as Apache Airflow are pivotal in meeting these needs, as they effectively support the implementation of the Workflows as Code paradigm. Nevertheless, despite its considerable advantages, developers still face challenges due to the specialized demands of workflow orchestration and the complexities of distributed execution environments. In this paper, we manually study 1,000 sampled Stack Overflow posts derived from 9,591 Airflow-related questions to understand developers' challenges and root causes while implementing Workflows as Code. Our analysis results in a hierarchical taxonomy of Airflow-related challenges that contains 7 high-level categories and 14 sub-categories. We find that the most significant obstacles for developers arise when defining and executing their workflow. Our in-depth analysis identifies 10 root causes behind the challenges, including incorrect workflow configuration, complex environmental setup, and a lack of basic knowledge about Airflow and the external systems that it interacts with. Additionally, our analysis of references shared within the collected posts reveals that beyond the frequently cited Airflow documentation, documentation from external systems and third-party providers is also commonly referenced to address Airflow-related challenges. 


\end{abstract}
\begin{keyword}
Workflows as Code \sep Apache Airflow \sep Workflow Orchestration \sep Stack Overflow \sep Empirical Study
\end{keyword}




\maketitle
\section{Introduction} \label{sec:introduction}
The advent of large-scale data has revolutionized the design and development of software systems. New software systems, characterized by their intensive data processing capabilities, are increasingly pivotal across various sectors, including healthcare, finance, manufacturing, and the automotive industry, among others~\citep{kersting2020se4ml}. We refer to these systems as \textit{data-intensive software systems}. 

Data-intensive software systems often involve executing a series of interconnected activities, collectively forming a \textit{workflow}. These workflows are complex, encompassing various components like code, data from multiple sources, specific configurations (such as execution conditions), and the necessary infrastructures~\citep{sculley2015hidden, munappy2019data}. The specific activities within these workflows vary depending on the application context. For instance, an ETL (Extract, Transform, Load) workflow typically involves activities that begin with extracting data from various sources. This data is then transformed into a usable format before being loaded to its final destination for direct user access or subsequent processing tasks. In the context of Machine Learning (ML)-enabled software systems, workflows typically include activities such as loading data from various sources, preprocessing the data to prepare it for analysis, engineering features, executing machine learning algorithms, validating, deploying the solution, and continuously
monitoring its performance~\citep{amershi2019software, ml}.

The manual development and management of these workflows tends to be both time-consuming and prone to errors. This is largely due to their reliance on complex execution conditions, the handling of large-scale data, and the necessity for iterative development and experimentation~\citep{munappy2020data,sculley2014machine}. For instance, while some workflows are designed for daily execution, others are activated by specific triggers, such as the arrival of new data. This diversity requires not just accurate scheduling, but also adaptable systems that can smoothly incorporate new data. The challenges are manifold, encompassing the potential for execution errors, difficulties in maintaining the accuracy of large datasets, and the need for continuous adaptation to optimize performance. 


In response to these challenges, the paradigm of ``\textit{Workflows as Code}'' has emerged as a programmable approach, enabling orchestration (defining, scheduling, and monitoring) of workflows using code. This approach integrates software engineering principles into workflow development and management, offering several advantages over traditional methods. By treating workflows as code, organizations can leverage automation and code-driven management, enhancing the development, update, and integration processes of workflows with other tools and platforms. As such, there has been a notable shift towards embracing the Workflows as Code paradigm in recent years. This has led to a new generation of orchestration platforms such as Apache Airflow\footnote{\url{https://airflow.apache.org/}}, 
Luigi\footnote{\url{https://github.com/spotify/luigi}}, and Dagster\footnote{\url{https://dagster.io/}}. Developers have increasingly adopted these platforms to implement Workflows as Code, empowering them to streamline operations and enhance collaboration across teams.


While the Workflows as Code paradigm has enabled efficient workflow development, it also introduces new challenges. The flexibility in coding methods can lead to workflows that are difficult to launch or debug, compounded by the lack of established best practices for development. Moreover, different from traditional systems, developers might require a steep learning curve to understand the concept of workflow development and management~\citep{ApacheAirflowIBMBlog}. 
Additionally, in complex data processing scenarios involving numerous datasets, transformations, and conditional logic, it can be hard to see how data flows and what causes problems or delays. Despite these potential challenges, no prior research has investigated the implementation of Workflows as Code in software development.

To fill this knowledge gap, this paper presents an
empirical study to understand the types and root causes of challenges developers face while implementing Workflows as Code leveraging workflow orchestration platforms. 
The result of such a study will help developers quickly understand the fundamental difficulties and common pitfalls associated with Workflows as Code. For our study, we focus on Airflow, which has recently risen to prominence as a premier open-source orchestration platform, significantly simplifying both the development and management of workflows in data-intensive software systems~\citep{GitHubStar}.
World-leading corporations, including \textit{Adobe}, \textit{Astronomer}, \textit{Etsy}, \textit{Google}, \textit{ING}, \textit{Paypal}, \textit{Qubole}, \textit{Quizlet}, \textit{Reddit}, \textit{Square}, and \textit{Twitter}, have embraced Airflow for managing intricate software tasks, especially those workflows in their data-intensive systems~\citep{apacheBlog, surveyAirflow}.


In line with common practices of using question-and-answer sites, such as Stack Overflow (SO), to explore the challenges developers face in specific software engineering tasks~\citep{yang2016security,rosen2016mobile, ahmed2018concurrency, alshangiti2019developing, bagherzadeh2019going,wang2022empirical}, we gathered a dataset of 9,591 Airflow-related questions and their corresponding answers from SO. Utilizing this data, we conduct an empirical investigation to address three research questions:

\begin{itemize}
    \item \textbf{RQ1: \rqone} We sampled a subset of 1,000 Airflow-related questions and identified a comprehensive hierarchical taxonomy of challenges faced by developers, encompassing 7 high-level categories and 14 subcategories. The most common inquiries revolve around seeking assistance in defining workflows and tackling problems encountered during workflow execution. Developers also face other difficulties, such as setting up deployment and production environments for Airflow and ensuring the quality of workflows.  
    \item \textbf{RQ2: \rqtwo} We examined each post considered in RQ1 along with corresponding accepted or highest-voted answers and comments, and identified the root causes of the identified challenges. We identified 10 types of root causes underlying the challenges that developers encounter with Airflow. These root causes highlight a two-fold issue: on the one hand, Airflow's official documentation does not sufficiently guide users; on the other hand, there is a lack of developers' understanding and knowledge regarding Airflow, its interactions with external systems, as well as broader aspects of programming and DevOps practices.

    \item \textbf{RQ3: \rqthree} We extracted 16,059 references of URLs shared by developers within the 9,591 collected posts. We observed that the most frequently referenced resource is Airflow's official documentation. However, our analysis also shows a significant reliance on documentation from external systems and third-party resources. This trend indicates the diverse range of information sources developers consult to address Airflow-related challenges.
 
\end{itemize}

Our research uncovers the types of common problems in implementing Workflows as Code. More specifically, this paper makes the following main contributions:
\begin{itemize}
    \item We are the first to explore the challenges faced by developers when implementing Workflows as Code using Apache Airflow. We propose two taxonomies to characterize these challenges and their root causes. We also discuss their prevalence. 
    \item We explore how online documentation resources with different relevance to Airflow are referred to via shared links and present our observations. 
    \item We discuss the implications of our findings from both research and practice perspectives and provide a replication dataset \footnote{\url{https://github.com/RISElabQueens/dataset_developers_challenges_WaC}} with annotated SO posts at different levels from various aspects (challenge, root cause, references) to foster future research.
\end{itemize}

The rest of the paper is organized as follows: \cref{sec:background} discusses the background of Airflow and related work. \cref{sec:data_collection} presents the data collection overview and \cref{sec:preliminary_study} performs the preliminary studies. \cref{sec:empirical_inv} details the three research questions. \cref{sec:discussion} presents discussions on our findings. \cref{sec:threats} discusses the threats to validity and finally, \cref{sec:conclusion} concludes the paper.

\section{Background and related work} \label{sec:background}
\subsection{Background}


\subsubsection{Platforms supporting Workflows as Code}
With the rise of data-intensive products and decision-making, open-source software engineering platforms embracing the Workflows as Code paradigm have become extremely popular in the last five years.


Airflow, Prefect\footnote{\url{https://www.prefect.io/}}, Dagster\footnote{\url{https://dagster.io/}}, Luigi\footnote{\url{https://github.com/spotify/luigi}}, Metaflow\footnote{\url{https://metaflow.org}} and Flyte\footnote{\url{https://flyte.org/}} enable workflow definitions in Python. 
Prefect and Dagster, supported by their cloud offerings, Prefect Cloud and Dagster Cloud allow practitioners to define end-to-end data pipelines and ML pipelines.
Metaflow, developed by Netflix, can be used to define DAG-based data and ML pipelines. It provides versioning of artifacts, reproducibility, and native support for experimentation~\citep{berg2019open}. 
Motivated by the ``model card'' concept~\citep{mitchell2019model}, researchers have developed a prototype tool on top of Metaflow that generates documentation of ML pipelines termed as ``DAG card''~\citep{tagliabue2021dag}.
Flyte, developed by Lyft, is a pipeline automation platform for complex data and ML pipelines at a scale where pipelines can be defined using their REST/gRPC API with SDK support for Java, Python and Scala.

Argo\footnote{\url{https://argoproj.github.io/workflows/}}, Azkaban\footnote{\url{https://azkaban.github.io/}}, and Pachyderm\footnote{\url{https://www.pachyderm.com/}} enable workflows to be defined via JSON, YAML, or XML files. 
Argo natively runs parallel jobs on Kubernetes, an open-source container orchestration system that automates software deployment, scaling, and management. 
Azkaban is a batch workflow job scheduler created at LinkedIn to run Hadoop jobs.
Data-driven pipelines are defined using Pachyderm, where data changes initiate the trigger. It provides data lineage with data versioning, and its usage ranges from complex data processing systems and ML system development to research in biotech and life sciences~\citep{novella2019container}.
These platforms differ in their supported code formats, core functionalities, integration capabilities, and target audience.




\begin{figure*}[ht]
    \centering
    \includegraphics[width=0.9\linewidth]{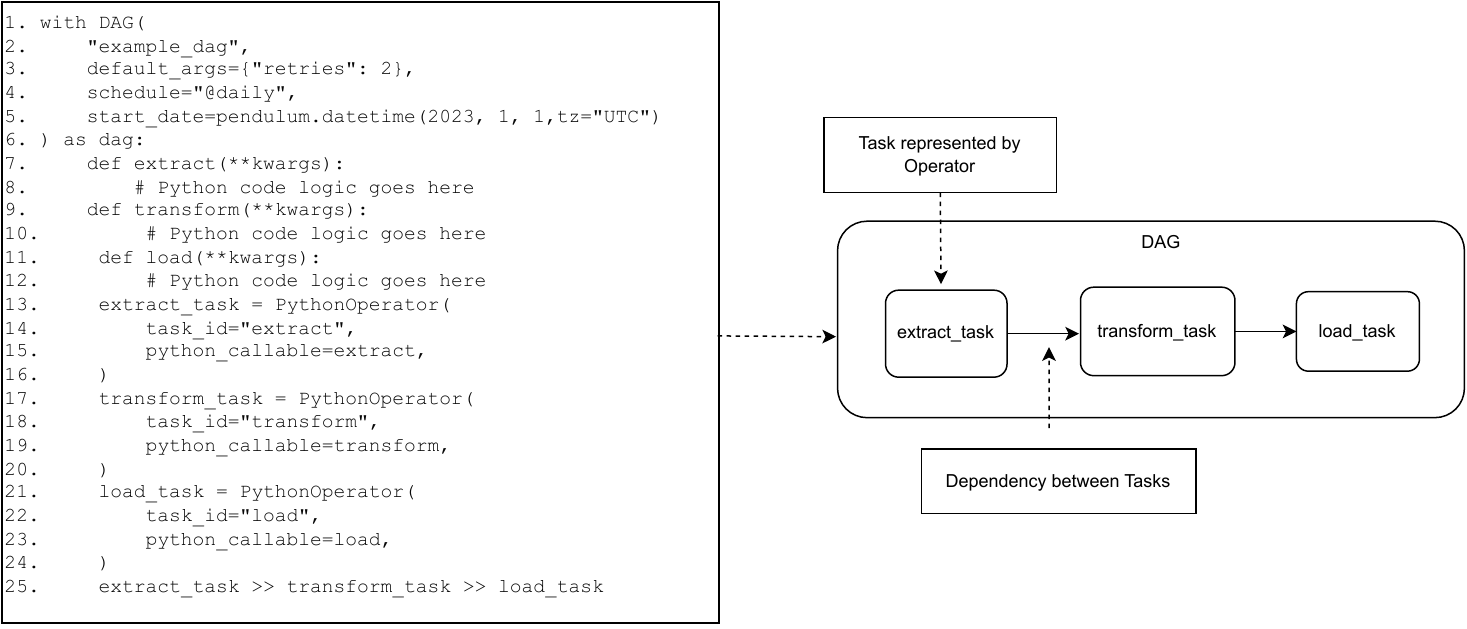}
    \caption{``Workflows as Code'' through Directed Acyclic Graphs (DAGs), visually capturing the seamless orchestration of tasks and dependencies in a concise, efficient manner.}
    \label{fig:example}
\end{figure*}

In our research, we have chosen Apache Airflow from among the platforms that support Workflows as Code. This decision was made based on Airflow’s widespread adoption within the industry and its popularity on Stack Overflow. A detailed discussion is elaborated in \cref{sec:threats}.

\subsubsection{Airflow}
Airflow, initially developed by Airbnb in 2014, has become a widely adopted open-source workflow orchestration tool for the development, scheduling, and monitoring of workflows. It entered the Apache incubator in 2016 and achieved the status of a top-level project under the Apache Software Foundation in 2019~\citep{apacheBlog}. Renowned for its proficiency in managing data-intensive workflows, Airflow has found extensive application in areas such as ETL processes, where the complexity of workflows necessitates robust management and automation.
In the rest of this subsection, we describe how a workflow is defined using \Airflow and the architecture and ecosystem of the platform.

\vspace{0.1cm}
\noindent \textbf{Workflows as Code using Airflow.} Airflow positions itself as a platform embodying the Workflows as Code paradigm~\footnote{\url{https://airflow.apache.org/docs/apache-airflow/stable/index.html}}. In Airflow, workflows are constructed using Python scripts, which outline \textit{Directed Acyclic Graphs (DAGs)}. These graphs are ``directed'' - indicating a predetermined order of task execution, and ``acyclic'' - guaranteeing the absence of cycles or loops within the workflow. In each DAG script, the tasks are represented as nodes, with the dependencies between these tasks depicted as edges. The most commonly used approach for implementing tasks is through \textit{operators}, as they offer a versatile and straightforward method for defining various types of tasks. Operators encapsulate the logic necessary to perform specific actions, such as running a Python function, executing a Bash command, transferring data between systems, or interacting with external services like databases or cloud platforms. Besides operators, Airflow also offers alternative approaches such as the TaskFlow API~\citep{TaskFlowAPI} or custom operator implementation for more specialized requirements.

Figure~\ref{fig:example}(left) illustrates an example of a workflow defined in Airflow using a DAG named ``\textit{example\_dag}'' (Line 2). This DAG has three parameters (Line 3-5). The ``\textit{default\_args}'' parameter is used to set common parameters for the DAG, such as defining the number of retries as 2. The ``\textit{schedule}'' parameter and the `` \textit{start\_date}'' parameter determine that this DAG is scheduled to execute daily, commencing from January 1, 2023, in Coordinated Universal Time (UTC).

The sample workflow shown in Figure~\ref{fig:example} (left) consists of three tasks: ``\textit{extract\_task}'', ``\textit{transform\_task}'' and ``\textit{load\_task}'' (Line 13-24). These tasks are responsible for the essential operations in ETL workflow: data extraction, transformation, and loading, respectively. Each task is based on a Python function, i.e., extract, transform, and load(Line 7-12). Each task is executed via a PythonOperator~\citep{PythonOperator}, a built-in operator provided by Airflow to execute Python callables. The Python functions act as placeholders for actual Python code, which will be executed during the respective task runs in the DAG. For simplicity of the example, we omit the detailed source code that developers should write to implement the corresponding logic. The three tasks are structured in a linear sequence (Line 25), ensuring that the output of one task is the input for the next. 

Figure~\ref{fig:example} (right) shows an abstract of the sample workflow. This graphical representation (a simple linear relationship in this case) showcases the sequence of tasks and their dependencies. 
In practice, the dependencies between tasks in a workflow can indeed become more complex than what is depicted in the abstract representation shown in Figure~\ref{fig:example} (right). 
For instance, tasks may have conditional dependencies, where their execution depends on the outcome of other tasks or external factors. Additionally, tasks may have parallel dependencies, where multiple tasks need to be completed before proceeding with subsequent tasks.

\noindent\textbf{Airflow Components.} Airflow's architecture is composed of a \textit{scheduler}, \textit{web server}, \textit{metadata database}, \textit{executor}, and \textit{workers}. We briefly discuss each element below:

\begin{itemize}
    \item \textit{Metadata Database}: Airflow uses a metadata database to store information about DAGs, tasks, and their status. This allows for tracking and managing the state of workflows.

    \item \textit{Web Server}: Airflow provides a web-based user interface to monitor and manage workflows. Users can visualize DAGs, view task logs, and manually trigger or pause DAG runs.
    
    \item \textit{Scheduler}: The scheduler monitors all tasks and DAGs to decide what needs to be run, based on dependencies and schedules. It sends tasks to be executed to the executor. Different types of executors are available in Airflow
including the LocalExecutor, which handles tasks
locally for parallel processing, and the SequentialExecutor, ideal for sequential task execution in simpler
setups. For distributed task execution, Airflow provides the CeleryExecutor, utilizing Celery for task
distribution across multiple machines, and the KubernetesExecutor, which scales tasks by creating separate
pods within a Kubernetes cluster. In environments designed for distributed and parallel execution, tasks are allocated across various workers, i.e., the computational units responsible for executing tasks, enabling them to be executed simultaneously.

    \item \textit{A folder of DAG files:} DAGs are typically stored as Python files in a designated folder. The scheduler monitors this folder, reading the DAG files to understand the task dependencies and execution schedules, and determine what tasks to run and when to run them, thus ensuring the smooth execution of workflows.
    
\end{itemize}


\noindent \textbf{The Airflow Ecosystem.}
The Airflow ecosystem has evolved to support fully managed workflow orchestration services such as Google Cloud Composer and Amazon Managed Workflows for Apache Airflow (MWAA). Google Cloud Composer offers a managed Apache Airflow service on Google Cloud Platform (GCP), facilitating seamless workflow orchestration with native GCP integrations and automatic updates. Similarly, MWAA provides a managed Apache Airflow environment on AWS, offering features like auto-scaling and integration with AWS services. Astro, provided by Astronomer, offers a modern data orchestration platform powered by Apache Airflow, enabling data engineers, scientists, and analysts to build, run, and observe Workflows as Code. Azure Data Factory Managed Airflow and Yandex Managed Service for Apache Airflow offer similar managed services on Azure and Yandex Cloud platforms, respectively. Additionally, solutions like Airflow with Restack and DoubleCloud Managed Service for Apache Airflow provide flexibility by enabling users to deploy Apache Airflow on their preferred cloud infrastructure or managed platforms.


\subsection{Related Work}

\subsubsection{Challenges on workflow development and management}
To the best of our knowledge, there has been no prior research investigating the challenges and implementation practices of Workflows as Code. The closest related studies focus on the difficulties encountered in the manual development and management of workflows within data-intensive systems.

Munappy et al.~\citeyear{munappy2020data} interviewed developers from the telecommunication and automobile domains and revealed three primary challenges in data pipeline management: infrastructure, data quality, and organizational issues. The authors highlighted the critical need for standardization across data pipelines, enhanced traceability, and the adoption of DataOps practices to address these challenges. Another work from the same authors~\citep{munappy2019data} conducted semi-structured interviews with seven deep learning (DL) experts. They identified the data management challenges in DL systems across seven stages, i.e., data collection, exploration, preprocessing, dataset preparation, data testing, deployment, and post-deployment. 
Polyzotis et al.~\citeyear{polyzotis2018data} identified fundamental challenges in ML data management that are applicable across various ML platforms, based on insights gained from constructing data management infrastructure for TFX. They highlight three main challenges: understanding the data, validating and cleaning it, and preparing data for ML tasks.



The existing reported challenges in manual workflow development and management underscore the need for adopting Workflows as Code. This programmable approach could facilitate traceability and standardization across data and machine learning pipelines. 

\subsubsection{Studies on developers' faced challenges leveraging Stack Overflow}
Stack Overflow (SO) serves as a valuable platform for understanding software engineering practices from developers' perspectives. Researchers have conducted empirical studies leveraging SO, and these studies typically fall into two categories: studies that analyze the general types of questions asked by developers on SO and those that delve into domain-specific challenges reflected in the questions.

Previous research ~\citep{allamanis2013and, beyer2020kind, cummaudo2020interpreting} has identified general types of inquiry on SO, including conceptual questions, how-to guides, API usage, error troubleshooting, and discrepancies, shedding light on the prevalent topics of discussion. However, although these taxonomies provide a useful framework for understanding general programming queries, they do not capture the nuanced challenges developers face within specialized domains.

Domain-specific research has been conducted across various fields, leading to the creation of taxonomies describing challenges and underlying root causes in specific application domains. For instance, an analysis of 2,758 questions related to Puppet identified 16 challenge categories specific to configuration as code~\citealp{rahman2018questions}. Similarly, focused studies have been conducted in numerous domains, including Apache Spark~\citep{wang2022empirical}, machine learning~\citep{alshangiti2019developing}, big data~\citep{bagherzadeh2019going}, mobile development, deep learning development~\citep{zhang2019empirical} and deep learning deployment~\citep{chen2020comprehensive}, each revealing unique challenges pertinent to their respective areas.

In \cref{sec:discussion}, we compare our taxonomies to the empirical studies that examine the challenges developers encounter by mining questions on SO.

\section{Data Collection}\label{sec:data_collection}

To gather posts from Stack Overflow (SO), we utilized the Stack Exchange Data Explorer web interface\footnote{\url{https://data.stackexchange.com/stackoverflow/queries}}. This tool facilitates the retrieval of up-to-date SO posts based on specific criteria, such as tags related to our research interest. We opted not to use the SOTorrent dataset~\citep{baltes2018sotorrent}, a commonly used SO data dump, because it has not been updated since December 2020. Figure~\ref{fig:study_design} provides an overview of our data collection process. Our data collection methodology, following previous studies~\citep{ponzanelli2014understanding, rosen2016mobile}, involves four steps:

\begin{figure*}[t]
    \centering
    \includegraphics[width=0.98\linewidth]{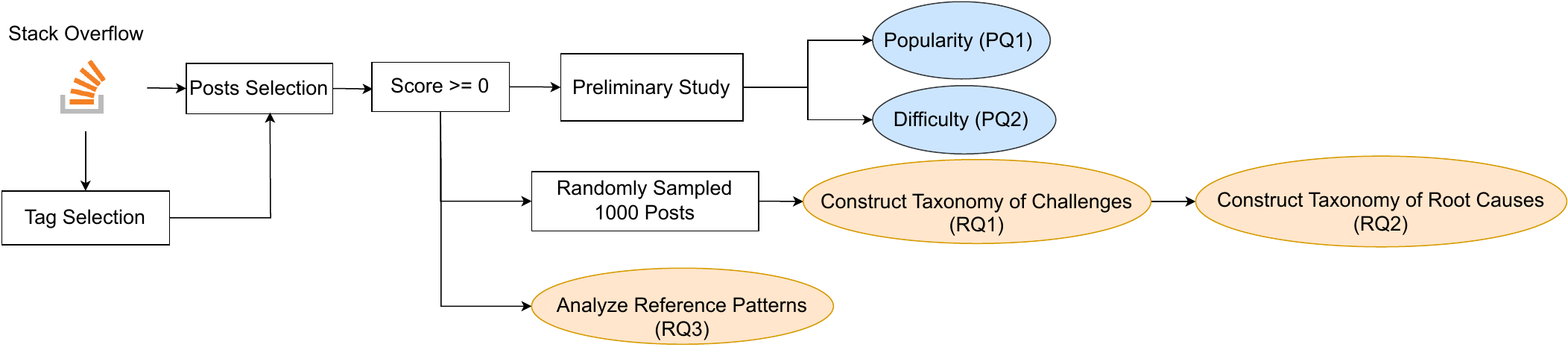}
    \caption{Overview of our data collection process.}
    \label{fig:study_design}
\end{figure*}

\textbf{Step 1: Identify Airflow-related tags.} We manually compiled a list of tags containing the keyword 'airflow', including \textit{airflow}, \textit{airflow-scheduler}, \textit{airflow-2.x}, \textit{airflow-api}, \textit{airflow-taskflow}, \textit{airflow-webserver}, \textit{airflow-connections}, \textit{airflow-xcom}, \textit{airflow-celery}, and \textit{airflow-k8s}.

\textbf{Step 2: Download Airflow-related question posts.} Using the Stack Exchange Data Explorer, we retrieved all questions (totaling 9,737) tagged with any of the 10 identified tags. These posts date from September 03, 2015, to April 30, 2023, and include metadata such as the post identifier, type, creation date, title, body, tags, view count, score, favorite count, and the identifier of the accepted answer (for questions).

\textbf{Step 3: Filter out negative questions.} We eliminated questions with a negative score, resulting in a total of 9,591 questions.

\textbf{Step 4: Collect contextual information.} For each retained question post from Step 3, we gathered its answers and all comments associated with both the questions and answers.

In the end, our dataset comprises 9,591 SO questions, 10,386 answers, and 21,298 comments. This dataset supports our preliminary study and analytics for the three research questions (RQs). For the preliminary analysis, we utilized all questions (9,591) and corresponding answers (10,386) from our dataset to examine the popularity and difficulty of Airflow-related questions. For RQ1 and RQ2, we randomly sampled 1,000 question posts from our dataset, along with their answers. For RQ3, we developed a script to automatically extract all 16,059 shared URLs from the collected data, including questions, answers, and comments. The detailed approach for each RQ will be presented in the next sections.




\section{Preliminary Study}\label{sec:preliminary_study}
We conduct a preliminary quantitative analysis on the data (9,591 questions and 10,386 answers) collected in \cref{sec:data_collection}.  
This preliminary study aims to answer two questions.

\textbf{\pqone}

\noindent\textbf{Motivation:} Analyzing the popularity of Apache Airflow-related questions on SO is critical for understanding Airflow's prominence and impact within the developer community. Through a quantitative examination of Airflow-related questions, we can ascertain the degree of interest and engagement among developers and how it has evolved over the past years, which in turn reveals Airflow's traction and its potential establishment as a key tool in automated workflow development. 

\vspace{0.1cm}
\noindent\textbf{Approach:} Following prior studies~\citep{alshangiti2019developing, chen2020comprehensive}, we assess the popularity of Airflow-related questions by examining both the volume of relevant questions on SO and the number of SO users posing these questions annually, spanning from 2015 to 2022. We exclude the year 2023 from this analysis, as our data collection only extends up to April of that year.

\begin{figure}[h]
    \centering
    \includegraphics[width=0.75\linewidth]{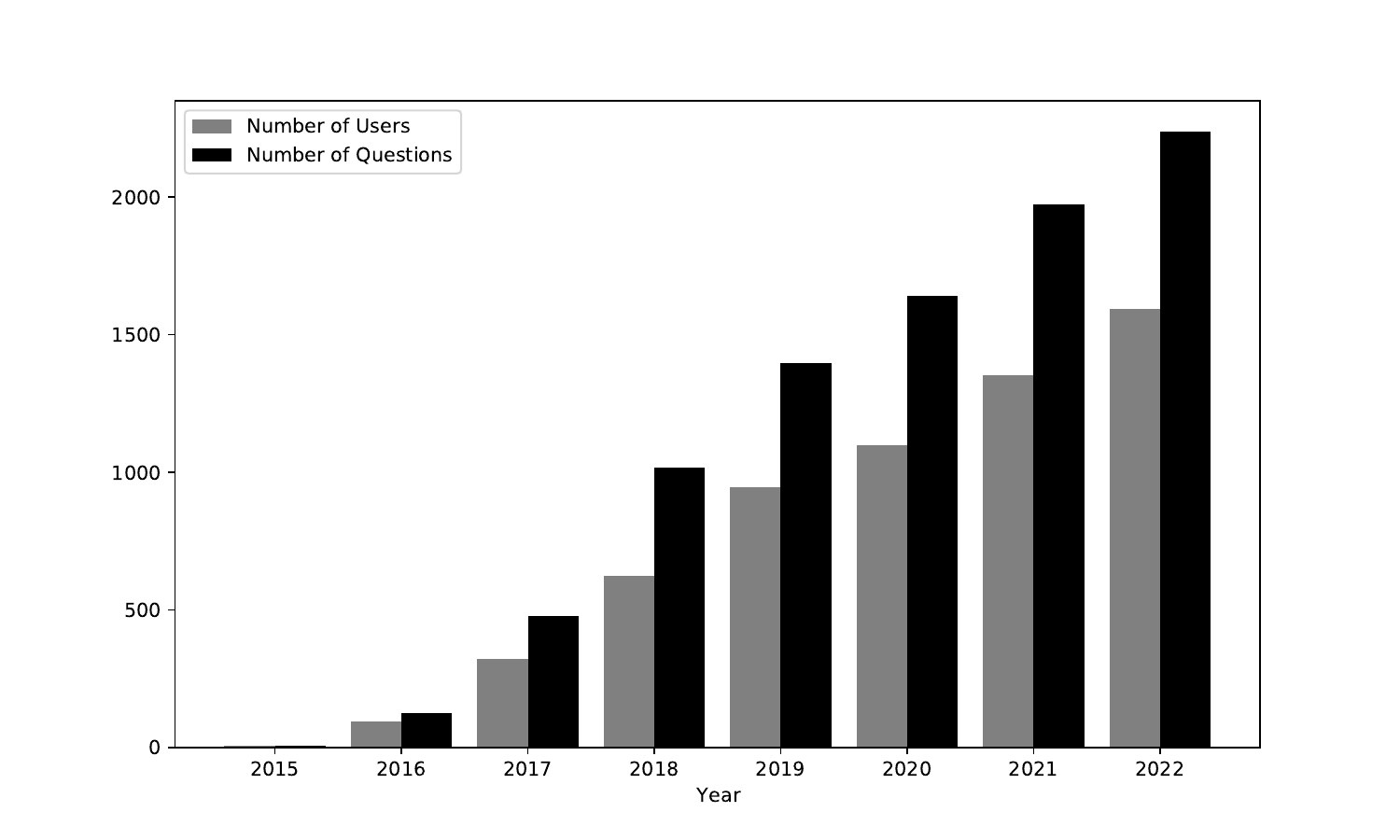}
    \caption{The upward trajectory of the number of questions and users asking about Airflow over the years.}
    \label{fig:line_plot}
\end{figure}

\vspace{0.1cm}
\noindent\textbf{Result:}
Figure~\ref{fig:line_plot} presents the annual trend in the popularity of Airflow-related questions, in terms of question number and involved user number. The result shows a significant upward trend in Airflow-related questions on SO from 2015 to 2022, signifying a steady escalation in user interest in Apache Airflow. Between 2019 and 2022, the annual increase in both users and questions on the platform ranged between approximately 16\% and 18\%. The results confirm the rising popularity and adoption of  Airflow among developers. 

\textbf{\pqtwo}

\noindent\textbf{Motivation:} Airflow is challenging for newcomers to onboard~\citep{ApacheAirflowIBMBlog}. In this question, we investigate the difficulty of answering Airflow-related questions, specifically by examining the time taken for these questions to receive a working solution. An answer to this question will not only quantify the complexity involved in resolving Airflow-related issues but also provide insights into the learning curve and potential knowledge gaps within the Airflow community. 

\noindent\textbf{Approach:} Similar to prior SO studies in different domains~\citep{yang2016security,rosen2016mobile, ahmed2018concurrency, alshangiti2019developing, bagherzadeh2019going}, we measure the difficulty of Airflow-related questions in two steps. We first collect Airflow-related questions with accepted answers and report the ratio of questions with accepted answers. For each selected question, we then calculate the duration between the posting time of the question and the accepted answer.

\noindent\textbf{Result:} Our data shows that only 36.4\% (3,491 out of 9,591) of Airflow-related questions on Stack Overflow have accepted answers. As indicated in Table~\ref{tab:difficulty}, this proportion is similar to that for deep learning development questions (37.3\%), but it is lower compared to questions in the domains of big data, concurrency, and mobile development.

\begin{table*}[t]
\centering
\caption{Difficulty of Airflow-related questions versus other domains (ranked by median time to receive accepted answers in decreasing order). The most challenging domain for each metric is emphasized.}
\resizebox{13.5cm}{!}{%
\begin{tabular}{p{0.70\linewidth}p{0.40\linewidth}p{0.80\linewidth}}
\hline
\toprule
    \textbf{Domain} & \textbf{\% of accepted answers} & \textbf{Median time to receive accepted answers (in minutes)}
    \\                                                                          \midrule
Airflow & 36.4\% & \textbf{527}
 \\   
 \midrule

Deep Learning deployment (\cite{chen2020comprehensive})  & \textbf{29.3}\% & 404.9 \\                                                                                 
\midrule
Deep Learning development (\cite{chen2020comprehensive}) & 37.3\% & 145.8  \\                                                                                                                              
\midrule
     Big Data (\cite{bagherzadeh2019going}) & 39.5\% & 198 
      \\                                                                        
 \midrule
Mobile (\cite{rosen2016mobile}) & 45\% & 55
 \\  
    \midrule
Concurrency (\cite{ahmed2018concurrency}) & 51.2\% & 42
 \\       
 \bottomrule
\label{tab:difficulty}
\end{tabular}}
\end{table*}

Regarding the median response time, it takes 527 minutes to receive an accepted answer for an Airflow-related question. This duration, as outlined in Table~\ref{tab:difficulty}, is notably longer than the response times for other specialized domains explored. For example, mobile development questions typically receive an accepted answer within just 55 minutes. Table~\ref{tab:distribution_difficulty} details the distribution of accepted answers over various time frames within a subset of 3,478 posts. Notably, only 1.5\% of accepted answers arrive within 5 minutes, 20.3\% within an hour, and 85.7\% within a week. 

\begin{table*}[htbp]
  \centering
  \caption{The percentage of questions (3,478 with accepted answers) that receive an accepted answer within a specified time.}
  \resizebox{9cm}{!}{%
  \begin{tabular}{p{0.70\linewidth}p{0.40\linewidth}}
    \toprule
    \textbf{Accepted Answer within} & \textbf{Out of 3,478 questions} \\
    \midrule
    5 minutes & 1.5\% \\
    \midrule
    1 hour & 20.3\% \\
    \midrule
    1 day & 65.6\% \\
    \midrule
    7 days & 85.7\% \\
    \midrule
    1 year & 98.6\% \\
    \bottomrule
    \label{tab:distribution_difficulty}
  \end{tabular}}
\end{table*}


\begin{summary_PQ}{}{}
While the popularity of Airflow-related questions on Stack Overflow is on the rise, the significant median response time of 527 minutes for accepted answers suggests difficulties developers encounter in automated workflow development. This underlines the relevance and importance of analyzing the specific challenges faced by developers when using Airflow.
\end{summary_PQ}

\section{Empirical Study and Results}\label{sec:empirical_inv}







\subsection{\textbf{RQ1: \rqone}}

\noindent \textbf{Motivation:} The objective of RQ1 is to develop a taxonomy that summarizes the challenges developers encounter when implementing Workflows as Code with Airflow, as reflected in their questions on SO. This taxonomy offers a systematic framework for practitioners and researchers to understand and navigate the complexities associated with Airflow-related challenges. Additionally, by examining the prevalence of various challenge types, we aim to identify the most common issues newcomers should be aware of when employing Airflow in workflow development and management.

\subsubsection{Approach}

We employed open coding~\citep{seaman1999qualitative} for the manual categorization of challenge types, analyzing both the titles and bodies of selected Stack Overflow questions, including their embedded code snippets. Specifically, we randomly selected 1,000 questions from our dataset of \TotalQuestionPostsScoreGteZero questions (as detailed in \cref{sec:data_collection}). This sample size was chosen to ensure a 95\% confidence level with approximately a 3\% confidence interval, aligning with methodologies used in previous research~\citep{wang2022empirical}. The categorization process was conducted in two stages and required 344 man-hours to complete:

\begin{itemize}
    \item \textbf{Stage 1: Pilot construction.} Two co-authors began by independently categorizing 300 (30\%) randomly chosen questions from the sample. They carefully reviewed each question together, labelling them based on the mentioned functionality of Airflow highlighted in the post. Similar labels were merged into broader categories; for instance, ``\textit{automated triggering}'' and ``\textit{manual triggering}'' were merged under ``\textit{scheduling}''. Questions not relevant or focused on Airflow (e.g., a developer is asking for a method to update the \code{gcloud} tool on \code{Google Cloud Composer}(GCP) worker nodes clarifying that it is a GCP issue rather than Airflow\footnote{\url{https://stackoverflow.com/questions/63965159}}) were classified as ``\textit{irrelevant}''. Following open coding, our taxonomy, with its categories and subcategories, was developed in a bottom-up approach, grouping subcategories into main categories based on their relevance to a primary functionality. This iterative process involved continuous refinement of the taxonomy through repeated analysis of the questions and labels, ensuring each question was assigned to the most fitting challenge category.

    \item \textbf{Stage 2: Extension to the full set.} Next, the first two authors independently applied the established taxonomy from Stage 1 to categorize the remaining 700 questions. 
    The annotators achieved a Cohen's kappa coefficient of 0.81, indicating substantial inter-rater agreement~\citep{viera2005understanding,landis1977measurement}. They then discussed and resolved conflicts for the 700 questions, finalizing the taxonomy. 
  
\end{itemize}

\subsubsection{Results}

\begin{figure*}[ht]
    \centering
    \includegraphics[width=0.98\linewidth]{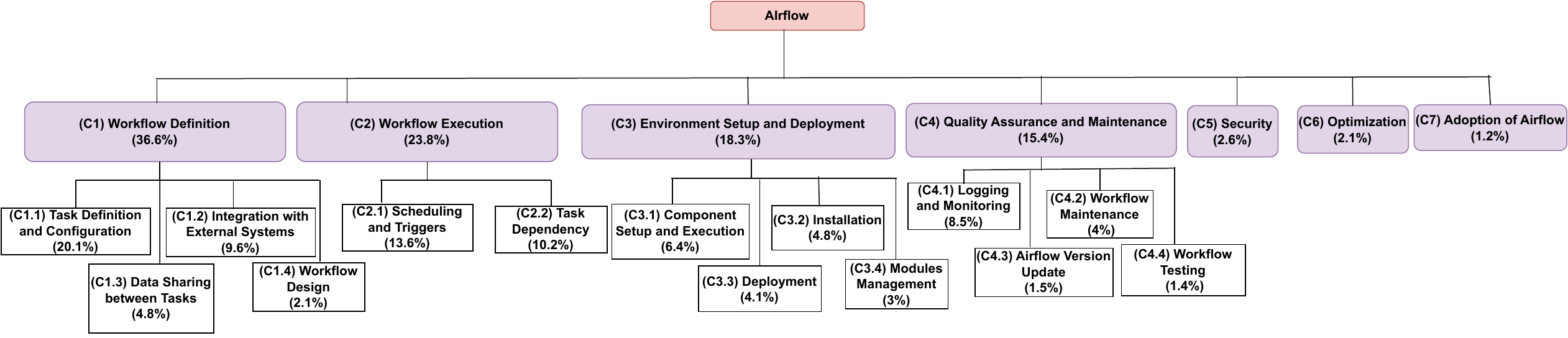}
    \caption{A hierarchical taxonomy of challenges faced by developers in utilizing Airflow.}
    \label{fig:rq1_result}
\end{figure*}

We observe that 58 out of 1000 (5.8\%) labeled questions are \textit{irrelevant} to Airflow. For the remaining questions, Figure~\ref{fig:rq1_result} presents an overview of our taxonomy, which is composed of 14 fine-grained subcategories and 7 high-level categories. Below, we describe each category in more detail.

\vspace{0.1cm}
\noindent \textbf{(C1) Workflow Definition}: Our analysis reveals that
\ChallengeWorkflowDefinition of the Airflow-related questions seek guidance in defining workflows. This finding aligns with expectations. Enabling ``\textit{Workflows as Code}'' is the core feature of Airflow. This paradigm, while powerful, presents a learning curve for developers, especially in comprehending how to translate complex workflow logic into executable code effectively. 


This high-level category is encompassed by the following four sub-categories, ordered by their prevalence:

\noindent \textbf{(C1.1) Task Definition and Configuration}: The most frequently encountered type of question in Category C1 asks about the defining and configuring tasks, accounting for \ChallengeTaskConfiguration of the Airflow-related questions. Tasks in Airflow are mostly defined using operators, which are specialized classes that encapsulate the logic for a specific task. Airflow offers a range of general operators like \code{BashOperator} and \code{PythonOperator} for executing a bash command or calling an arbitrary Python function, as well as provider-specific operators such as \code{BigQueryUpdateTableOperator} and \code{AWSAthenaOperator}, enabling integration with various services. Configuring a task thus involves selecting an appropriate operator and carefully setting a multitude of parameters associated with it, which can be a complex process for developers. For instance, developers often struggle with configuring the \code{BashOperator}, particularly when needing to set the \code{bash\_command} argument with dynamic values extracted from intricate \code{JSON} structures.\footnote{\url{https://stackoverflow.com/questions/74444283}}

In addition to using built-in operators, the creation of custom operators in Airflow is frequently necessary for workflows with specific demands. Yet, mimicking the comprehensive functionality of the built-in operators can be challenging. For example, a developer creating a custom operator might face difficulties with accessing the \code{task\_instance} object, a feature readily available in standard operators like \code{PythonOperator}, but not directly accessible in custom ones.\footnote{\url{https://stackoverflow.com/questions/59325584}}

\vspace{0.1cm}
\noindent \textbf{(C1.2) Integration with External Systems}: Our result reveals that \ChallengeIntegrationWithTheExternalSystems of the Airflow-related questions ask about issues when integrating external systems, such as data storage service and data processing frameworks, with Airflow. As a workflow orchestration tool, Airflow's ability to seamlessly interact with these external systems is crucial for the successful execution of tasks and workflows. Unfortunately, we observe that developers frequently face challenges in such integration process, starting from the very first step, i.e., establishing the connections between Airflow and external systems. For example, a developer had trouble while implementing a task using the \code{EmrCreateJobFlowOperator} in Airflow and is getting an error about an undefined connection ID (\code{emr\_default}).\footnote{\url{https://stackoverflow.com/questions/70222696}} This error suggests that the required connection information for Amazon EMR (a cloud big data platform) is missing. Besides connection issues, developers also encounter difficulties in the configuration of the external systems.



\vspace{0.1cm}
\noindent \textbf{(C1.3) Data Sharing between Tasks}: We observe that \ChallengeSharingDataBetweenTasks of the Airflow-related questions concerns about the exchange of data - information, results, or artifacts between various tasks within a workflow. Effective data sharing is crucial to the orchestration of complex data processing pipelines, yet developers often encounter obstacles in efficiently accessing or storing data between tasks. For example, one developer faced difficulties in devising a method to pass data to a subsequent task.\footnote{\url{https://stackoverflow.com/questions/69487557}} The task involved fetching data from ten REST API endpoints and saving each dataset into a data lake. The primary challenge for the developer was finding an optimal way to transfer large datasets from the data-fetching task to the data-saving task. The Airflow built-in method for data sharing, XCom\footnote{\url{https://airflow.apache.org/docs/apache-airflow/stable/core-concepts/xcoms.html}}, was not feasible due to its limitations on data size, motivating the developer to seek alternative solutions.

\vspace{0.1cm}
\noindent \textbf{(C1.4) Workflow Design}: In  \ChallengeDesignoftheWorkflow of the Airflow-related questions, developers did not ask specific technical details, but rather how to optimally design workflows tailored to specific scenarios. This is particularly challenging when workflows require intricate interactions between various services or systems. For instance, a developer asked how to design a workflow that downloads tables from a QA Redshift (an Amazon-managed data warehouse service) to an S3 bucket and then transfers them to another Redshift instance, using LocalStack (a mocked AWS service).\footnote{\url{https://stackoverflow.com/questions/71504505}}

\vspace{0.1cm}
\noindent \textbf{(C2) Workflow Execution}: Our analysis reveals that \ChallengeWorkflowExecution of the Airflow-related questions seek guidance in defining workflow execution order and addressing runtime issues during workflow execution. 
These issues can be intricate, stemming from diverse factors such as data dependencies, scheduling conflicts, or resource constraints. 

\vspace{0.1cm}
\noindent \textbf{(C2.1) Scheduling and Triggers}: A significant portion of Airflow-related queries, representing \ChallengeScheduling of the total, focuses on scheduling workflows and configuring triggers. These scheduling issues often originate from inaccuracies in defining schedule intervals or handling time zone variances. An example includes a developer encountering difficulties in precisely timing workflow executions.\footnote{\url{https://stackoverflow.com/questions/70228718}} In Airflow, triggers are conditions or events typically defined within a DAG (as shown in Lines 5 and 6 of Figure~\ref{fig:example}), enabling the automatic initiation of tasks under certain conditions. However, there are scenarios where developers need to trigger DAG, which runs externally and is facilitated through an HTTP request to the dagRuns endpoint of the Airflow REST API. This process can be challenging and error-prone, as illustrated by a specific case.\footnote{\url{https://stackoverflow.com/questions/67110383}} These findings reveal the need for more intuitive and user-friendly methods for scheduling and triggering in Airflow to enhance both automated and external execution processes.

\vspace{0.1cm}
\noindent \textbf{(C2.2) Task Dependency}: \ChallengeTaskDependency of Airflow-related questions is centred around task dependency issues. Task dependencies are the directed edges that determine how to move through the DAG, which defines a workflow. Properly managing task dependencies is crucial for ensuring that tasks are executed in the correct order. However, we observe that developers face difficulties in defining those dependencies, especially when encountering dynamic tasks (generated at runtime based on current data). For example, a developer was facing difficulties while trying to configure dependencies dynamically from tasks and their dependencies defined in a data frame.\footnote{\url{https://stackoverflow.com/questions/75166872}}

Moreover, effective task dependency management in Airflow extends beyond defining task relationships, and it also involves adapting to status changes during runtime. For example, if an upstream task fails to find files in an S3 (Amazon's Storage Service) bucket, it should trigger an email notification and stop a subsequent task, instead of executing the subsequent task. Developers often seek solutions for implementing such conditional execution flows within their workflows.\footnote{\url{https://stackoverflow.com/questions/68114312}}

\vspace{0.1cm}
\noindent \textbf{(C3) Environment Setup and Deployment}: Our analysis reveals that developers often encounter challenges when setting up Airflow in their development environment and deploying their DAGs in production, representing \ChallengeEnvSetup of all Airflow-related questions. This high-level category is encompassed by the following four sub-categories, ordered by their popularity.



\vspace{0.1cm}
\noindent \textbf{(C3.1) Component Setup and Execution}:
As described in \cref{sec:background}, Airflow is composed of four key architectural components: schedulers, a web server, a folder of DAG files, and a metadata database. For a successful launch of a DAG, these components (e.g., a MySQL database to serve as the metadata database) must be properly set up and correctly configured. Nonetheless, developers often encounter challenges in this setup process, as well as in executing those components. For instance, a developer encountered a ``disk I/O error'' while attempting to run the ``\textit{airflow scheduler}'' and ``\textit{airflow webserver}'' commands in Airflow.\footnote{\url{https://stackoverflow.com/questions/73311291}}

\vspace{0.1cm}
\noindent \textbf{(C3.2) Installation}: Our analysis shows that \ChallengeBasicSetup of Airflow-related questions report issues developers faced in the installation of Airflow. For example, a developer encountered a problem while attempting to install Airflow on a very specific environment, i.e., macOS Catalina with Python 3.8.2, using the ``\textit{pip install -airflow}'' command.\footnote{\url{https://stackoverflow.com/questions/64832243}} 

\vspace{0.1cm}
\noindent \textbf{(C3.3) Deployment}: We observe that \ChallengeDeployment of the Airflow-related questions are concerns about deploying Airflow from the development environment to the production environment. The primary reason for these concerns is the complexity involved in transitioning between different environments. This includes ensuring consistency in configurations, managing dependencies, and handling differences in infrastructure setups. Additionally, adapting workflows to the constraints and requirements of production environments, such as security protocols and resource limitations, can be a demanding task. These factors often require developers to make significant adjustments that are error-prone. For example, a developer experienced a deployment-related issue while trying to deploy Airflow on Azure Kubernetes Services (AKS) and load DAGs from a GitHub repository using Helm charts.\footnote{\url{https://stackoverflow.com/questions/71258379}}

\vspace{0.1cm}
\noindent \textbf{(C3.4) Module Management}: The least prevalent type of questions in C2 is concerned with the module management feature offered by Airflow, accounting for \ChallengeModuleManagement of Airflow-related questions. Airflow enables developers to import Python modules in DAG files for defining and executing tasks. These modules can encompass custom functions, classes, or additional code essential for task execution. However, configuring these dependent modules correctly could be challenging for developers. For instance, a developer experienced a \errorcode{ModuleNotFoundError}
when attempting to import a Python method from a file located in a local directory. 


\vspace{0.1cm}
\noindent \textbf{(C4) Quality Assurance and Maintenance:}
According to our analysis, \ChallengeWorkflowManagement of the Airflow-related questions concern the quality and ongoing maintenance of workflows managed through the platform. This high-level category contains the following four sub-categories, ordered by their popularity. 

\vspace{0.1cm}
\noindent \textbf{(C4.1) Logging and Monitoring}: In \ChallengeLoggingandMonitor of the Airflow-related questions, developers encountered challenges with Airflow's logging and monitoring capabilities. Given that workflows typically operate autonomously, effective observability through logging is crucial. Apache Airflow's logging framework is designed to offer comprehensive insights into task and DAG executions. While its default setup writes logs to the local file system—adequate for development and debugging phases—production environments, particularly in cloud settings, often necessitate remote logging capabilities to services like AWS S3, Google Cloud Storage, and Azure Blob Storage. However, developers frequently encounter hurdles in both local and remote logging setups. An example case is a developer who faced issues with the Airflow UI redirecting to localhost instead of the EC2 instance's IP address when trying to access task logs, requiring manual intervention to correct the URL.\footnote{\url{https://stackoverflow.com/questions/49784569}} Another developer struggled with configuring Airflow to write logs to an S3 bucket.\footnote{\url{https://stackoverflow.com/questions/63992194}}





\vspace{0.1cm}
\noindent \textbf{(C4.2) Workflow Maintenance}: Our analysis reveals that \ChallengeWorkflowMaintenance of the Airflow-related questions are related to challenges in workflow maintenance. The maintenance of workflows in Airflow presents several difficulties. One of the primary challenges is managing the dependencies and versioning of complex DAGs along with their associated libraries, particularly in the absence of a dedicated versioning mechanism for workflows within Airflow. An example from our annotated questions includes a developer seeking guidance on how to implement DAG versioning in Airflow, aiming to enhance the efficiency of management and maintenance processes.\footnote{\url{https://stackoverflow.com/questions/61796692}} Furthermore, as the number and complexity of tasks and workflows expand, scaling Airflow to adequately handle an increasing workload and meet escalating resource requirements emerges as another type of complexity.



\vspace{0.1cm}
\noindent \textbf{(C4.3) Airflow Version Update}: In some questions (\ChallengeMigration of the Airflow-related questions), developers reported issues faced when updating to newer versions of Airflow. Even with the support features provided by Airflow, such as automated methods for upgrading the metadata database, migration issues still occur. For instance, an error was found while a developer was migrating from Airflow version 1.10.15 to 2.2.5.\footnote{\url{https://stackoverflow.com/questions/72529432}} After the update, the developer encountered a problem when using the \code{BigQueryToGCSOperator} to transfer data from BigQuery to GCS. The process resulted in a \errorcode{NotFound}
exception and a 404 error, indicating that the job linked to the operator could not be found – a problem that did not exist in Airflow 1.10.15. The developer sought to understand the root cause of this compatibility issue and explore possible solutions for smooth migration.



\vspace{0.1cm}
\noindent \textbf{(C4.4) Workflow Testing}: Our analysis shows that developers sometimes (\ChallengeTesting of the Airflow-related questions) ask questions about testing of tasks and DAG in Airflow. Effective testing is essential for ensuring the reliability of workflows. This encompasses unit testing to evaluate individual components of a DAG, DAG testing to confirm the overall structure and integrity of the workflow, and integration testing to examine how the workflow interacts with external systems. Despite its importance, developers encounter challenges in successfully conducting these tests within the Airflow framework. For example, a developer faced difficulties while carrying out unit tests for an ETL process in Airflow, particularly aiming to ensure that their DAGs were free of cycles.\footnote{\url{https://stackoverflow.com/questions/67725703}}



%




\vspace{0.1cm}
\noindent \textbf{(C5) Security}: Our analysis reveals that \ChallengeSecurity of the Airflow-related questions focus on security implementation challenges within the platform. Airflow has a security model ensuring the integrity and confidentiality of the tasks being executed. The security model adopts robust authentication mechanisms, such as LDAP and OAuth. Despite these foundational security designs, developers often encounter complexities in setting up and managing these advanced security protocols. For instance, a developer encountered issues in the implementation of LDAP authentication in Airflow.\footnote{\url{https://stackoverflow.com/questions/71249044}} The specific challenge lies in safely managing the LDAP bind password within Airflow's configuration settings.


\vspace{0.1cm}
\noindent \textbf{(C6) Optimization}: Our analysis indicates that \ChallengeOptimization of the Airflow-related questions involves developers seeking advice on optimizing their Airflow configurations. Optimization in this context refers to enhancing the system's efficiency, responsiveness, and resource management to ensure workflows are executed both timely and smoothly. A common optimization challenge is to address excessive CPU usage in Airflow setups. For example, one user reported a high CPU utilization, up to 20\%, on an EC2 instance running Airflow with Docker Compose and LocalExecutor, even when the system was idle.\footnote{\url{https://stackoverflow.com/questions/67063149}} The user sought advice on configuration adjustments to lower CPU usage, as well as insights into the potential trade-offs of such changes.

\vspace{0.1cm}
\noindent \textbf{(C7) Adoption of Airflow}: We find that in \ChallengeAdoption of the Airflow-related questions, developers seek decision-making guidance on whether to adopt Airflow given their technical stack and requirements. 

\begin{summary_RQ1}{}{}
We identify a comprehensive hierarchical taxonomy for Airflow-related challenges faced by developers, encompassing 7 high-level categories and 14 subcategories. The most prevalent among these high-level categories encompasses questions that seek guidance in defining workflows and addressing issues that arise during workflow execution. Developers also face other difficulties, such as setting up deployment and production environments for Airflow and ensuring the quality of workflows. 
\end{summary_RQ1}




\subsection{\rqtwo}
\noindent\textbf{Motivation:}
In RQ1, we identified a wide spectrum of questions developers encounter when utilizing Airflow for their workflow development and management. However, similar types of questions can stem from different root causes, and conversely, diverse questions may originate from the same fundamental issue. Therefore, in RQ2, we perform a manual analysis to identify the underlying root causes of the  Airflow questions. By uncovering these root causes, our research can provide practitioners with valuable knowledge, allowing them to navigate common issues more effectively. Furthermore, the results can inspire future research into creating more efficient support mechanisms and tools for developers working with Airflow, thereby enhancing the overall user experience and effectiveness of the platform.

\subsubsection{Approach} 
We carried out a manual analysis on the same set of 1,000 questions previously sampled for RQ1, employing a similar two-stage open coding methodology. The coding process was completed over the course of approximately 450 hours.  

\begin{itemize}
    \item \textbf{Stage 1: Pilot construction.} In this stage, two co-authors collaboratively established the root causes for a pilot set of 300 posts. The 21 question posts previously categorized as ``irrelevant'' to Airflow during RQ1 were excluded from this analysis. To identify the root cause of each question, the annotators carefully examined each element of a question post, including its title, body text, code snippets, associated answers, comments, and even referenced URLs. This comprehensive review ensured that the root cause was identified based on its detailed explanation in the discussion. If the specific root cause for a question remained ambiguous, it was labeled as ``\textit{unclear}''. Specifically, 50 question posts were labeled as ``\textit{unclear}''.
    Through this process, the annotators developed a taxonomy encompassing 10 distinct root causes.
    
    \item \textbf{Stage 2: Extension to the full set.} Next, the two annotators individually applied the taxonomy established in Stage 1 to the remaining 700 questions, excluding 37 questions that are irrelevant to Airflow. The inter-rater agreement during the independent labeling is 0.79, measured by Cohen’s Kappa~\citep{viera2005understanding}, indicating substantial inter-rater agreement. The annotators then discussed and resolved all conflicts and finalized the taxonomy. In this stage, 128 question posts were labeled as ``\textit{unclear}''.
\end{itemize}


\begin{table*}[h]
\centering
\caption{
    Root causes derived from the manual analysis of 942 \Airflow-related posts in Stack Overflow.
  }  
\resizebox{12.5cm}{!}{%
\begin{tabular}{p{0.6\linewidth}p{0.77\linewidth}p{0.24\linewidth}}
\hline
\toprule
    \textbf{Root cause} & \textbf{Definition} & \textbf{\% of Questions} \\      \midrule
                                                                        
R1:\rcone
& 

Developers face many issues within Airflow implementations due to the omission or misconfiguration of settings related to workflow configuration at the task level and DAG level. 

& 18.9
    \\
    \midrule 
    R2:\rctwo
&
The wide range of configuration options and settings required for \Airflow, can pose challenges in ensuring optimal setup and performance.
 & 14

    \\
\midrule

    R3:\rcthree
 & Developers need to have a basic understanding of \Airflow's core concepts, including Directed Acyclic Graphs (DAGs), operators, and task dependencies, to effectively design, implement, and manage workflows.  & 13

 \\
 \midrule
 R4:\rcfour & Developers encounter challenges due to issues or defects, or unsupported features that require extra effort within the \Airflow platform. & 8.2

\\

\midrule 
    R5:\rcfive & Lack of knowledge about external systems/domains, including cloud storage, database systems, and managed Airflow services, can lead to several challenges. & 6.5

    \\
\midrule 
   

    R6:\rcseven
& When working with \Airflow, developers often encounter challenges due to missing dependencies, which can complicate both the setup of the necessary environment and the definition of their tasks. & 5.7
\\
\midrule 
    R7:\rceight
    & Developers encounter challenges due to knowledge gaps in Python and a lack of understanding of the development and DevOps process while working with \Airflow. & 5.6
    \\
\midrule 
 R8:\rcsix
&  Airflow provides a wide range of operators, each designed for different use cases and functionality. The selection of an operator that doesn't align well with the specific requirements of a task or job within a workflow. (Misuse/Unknown to operator) & 4.8
    \\

\midrule 
    
    R9:\rcnine
    & Complex workflows, with their intricate processes, dependencies and dynamic behavior present challenges for developers, making it difficult to manage and understand them effectively.
    & 3.1
    \\
\midrule 
    R10:\rcten
& Developers may use an outdated API that is no longer compatible with a newer version of Airflow. & 0.8
 \\

\midrule 
\midrule 

 Unclear
& Lack of information to derive a root cause. & 19.3
    \\

\\
\bottomrule
\label{tab:rootcause}
\end{tabular}}
\end{table*}

\subsubsection{Results} 
Table~\ref{tab:rootcause} presents the taxonomy of root causes we identified through our manual analysis of 942 Airflow-related questions (as mentioned in RQ1, 58 out of the 1000 sample questions are not relevant to Airflow). The findings indicate that the three most prevalent root causes are incorrect workflow configurations in DAGs, challenges with complex environment configuration, and a lack of basic knowledge about Airflow (e.g., concepts related to DAG). Combined, these three categories account for 46.7\% of all Airflow-related questions. Notably, 6.5\% of the challenges stem not directly from Airflow itself but from a deficiency in understanding the external systems with which Airflow interacts. Below, we describe each category in more detail.

\vspace{0.1cm}
\noindent \textbf{(R1) \rcone}: Misconfigurations in DAGs account for 18.9\% of the Airflow-related issues we analyzed. Such configuration challenges can arise at both the task and DAG levels, involving aspects like setting operator parameters for tasks or scheduling workflows. An example is a developer failing to correctly schedule a workflow involving a crucial but optional parameter named \textit{execution\_timeout}.\footnote{\url{https://stackoverflow.com/questions/54810074}}

We further investigated potential factors that may contribute to the prevalence of this category and found three. First and foremost, Airflow's documentation does not provide a centralized reference for all mandatory and optional parameters, leaving developers without clear guidance on their usage. Secondly, while Airflow allows for dynamic parameter configuration at runtime for tasks and DAGs, it lacks clarity on which parameters are compatible with this feature. Thus, developers often fail to configure dynamic parameters in their DAGs. Last but not least, Airflow's official documentation often omits critical information regarding the interdependencies and constraints of multiple parameters (e.g., two parameters should be configured together), leading to confusion and misconfiguration. For instance, understanding the interplay between scheduling parameters i.e., \code{schedule\_interval} and \code{start\_date} is needed to ensure the desired execution behavior of the DAG.\footnote{\url{https://stackoverflow.com/questions/74235924}}


\vspace{0.1cm}
\noindent \textbf{(R2) \rctwo}: We observe that 14\% of Airflow-related questions are rooted in complex environmental and component configuration. \Airflow is recognized for its versatile configuration options, supporting a wide array of infrastructures and use cases. These configurations are primarily managed via the ``airflow.cfg'' file, which offers extensive customization of settings. However, the platform's diversity in configurations can lead to significant challenges. For instance, one developer struggled with setting up the metadata database.\footnote{\url{https://stackoverflow.com/questions/70903197}} 

Challenges also arise when integrating Airflow with distributed task queueing systems like \code{CeleryExecutor}, especially when managing data storage access across multiple worker nodes.\footnote{\url{https://stackoverflow.com/questions/66750928}}

In containerized environments like Docker, or orchestration platforms like Kubernetes, configuring Airflow involves a deep understanding of these platforms' specific features, often going beyond the \textit{airflow.cfg} file to include Docker and Kubernetes settings. For instance, in Docker, the Dockerfile serves as the blueprint for building the Airflow image, which entails various steps: selecting a base image, installing dependencies, configuring Airflow, and adding DAGs and plugins. While Airflow provides a standard Dockerfile, it frequently requires customization. For instance, a developer encountered issues installing Python packages in a virtual environment due to a conflict with the PIP\_USER environment variable set in the original Dockerfile.\footnote{\url{https://stackoverflow.com/questions/73962053}} The default setting activated the --user flag, which was incompatible with the developer's attempts to install packages in the virtual environment, resulting in errors.

Configuration through the web interface can also initiate challenges. For example, despite configuring the connection details in Airflow's web interface, a developer encounters difficulties in making the connection work from within the Airflow container.\footnote{\url{https://stackoverflow.com/questions/70267145}}

\vspace{0.1cm}
\noindent \textbf{(R3) \rcthree}: Our analysis reveals that 13\% of the studied problems are related to a lack of basic knowledge of Airflow. Developing effective workflows in Airflow demands a solid grasp of its core concepts, including Directed Acyclic Graphs (DAGs), basic operators, task dependencies, and the Airflow Command Line Interface (CLI). Misunderstandings of these foundational aspects can lead to challenges.

An example involves a developer who faced difficulties with DAGs not appearing in the web UI when defined in separate files.\footnote{\url{https://stackoverflow.com/questions/70051020}} This issue stems from a misunderstanding of Airflow's DAG discovery mechanism, which relies on specific conventions to identify DAGs in Python files if DAGs are defined in multiple files. The developer's approach did not align with these conventions. To address this, another developer suggested including the term ``airflow'' in the file, even as a comment, to meet the discovery criteria. This issue exemplifies a common knowledge gap regarding the DAG discovery conventions in Airflow.


\vspace{0.1cm}
\noindent \textbf{(R4) \rcfour:} 8.2\% of the questions related to Airflow arise from inherent limitations in specific versions, such as bugs or unsupported features. These issues can significantly disrupt development, often requiring developers to devise workarounds or participate in the community's problem-solving efforts. For instance, a developer using Airflow 2.0.0 encountered an unexpected behavior where a workflow set for daily execution skipped a day in its schedule, despite being correctly set up and manually triggered.\footnote{\url{https://stackoverflow.com/questions/65897976}} This issue was traced back to a bug in the Airflow 2.0.0 release \footnote{\url{https://github.com/apache/airflow/issues/13434}}, which was subsequently fixed in the 2.0.1 update. In another case, a developer faced challenges in executing a dynamic number of tasks, with a maximum of four tasks in parallel.\footnote{\url{https://stackoverflow.com/questions/75192266}} This led to the introduction of a new feature in Airflow to better support dynamic task execution.\footnote{\url{https://github.com/apache/airflow/issues/29084}} Issues have also been reported with Airflow's Role-Based Access Control (RBAC) system, especially regarding its limited functionality when applied to developer actions through the Experimental API or the Airflow CLI.\footnote{\url{https://stackoverflow.com/questions/62759527}}

It is worth noting that bugs may occur not only in Airflow's core framework but also in provider-specific operators. For example, a bug introduced in version v7.0.0 of the Airflow Providers Google package could lead to task failures, necessitating workarounds such as reverting to an earlier version or using the BigQuery API directly.




\vspace{0.1cm}
\noindent \textbf{(R5) \rcfive}: Our analysis shows the root cause of 6.5\% of the studied problems is related to insufficient knowledge about external systems that Airflow interacts with through its tasks. For instance, a developer using the \code{GoogleCloudStorageComposeOperator} in Google Cloud Composer encountered the \errorcode{HTTP 429 Too Many Requests} error.\footnote{\url{https://stackoverflow.com/questions/69988105}} This error suggests the rate limit was exceeded for changes to the ``object path/file.csv'', likely due to Google Cloud Storage's write limit of once per second for the same object name. The issue indicates that the developer was unfamiliar with or overlooked these specific rate limits - a common concept that needs to be aware of when using Google Cloud Storage. In another scenario, a developer working with Amazon Managed Workflows for Airflow (MWAA) faced challenges as the platform restricted the installation of additional packages on the webserver \footnote{\url{https://stackoverflow.com/questions/68404064}}, which can be resolved if the developer knows how to establish connections to Google Cloud through the webserver.




\vspace{0.1cm}
\noindent \textbf{(R6) \rcsix}: Our analysis reveals that 5.7\% of the Airflow-related issues stem from developers making incorrect or sub-optimal choices of operators for their tasks. Airflow offers a range of built-in operators for basic tasks and provider-specific operators for more specialized tasks interacting with external systems. Each operator in Airflow is tailored for specific use cases, and choosing the wrong one can lead to inefficiencies or added complexities in the workflow. For example, a developer used the \code{PythonOperator} for tasks where the \code{PythonVirtualenvOperator} might be a more optimal choice, particularly for handling Python dependencies in a virtual environment.\footnote{\url{https://stackoverflow.com/questions/67615888}} 


Furthermore, navigating the broad selection of built-in and provider-specific operators can be challenging. In one instance, a developer sought advice on choosing the right operator to load a file from Google Cloud Storage (GCS) into BigQuery, knowing the operator for the reverse process but unclear about the best choice for this particular task.\footnote{\url{https://stackoverflow.com/questions/74191180}} The difficulty here lies in selecting from a diverse array of available operators \footnote{\url{https://airflow.apache.org/docs/apache-airflow-providers-google/stable/operators/cloud/bigquery.html}}, each suited to different aspects of interacting with GCS and BigQuery.


\vspace{0.1cm}
\noindent \textbf{(R7) \rceight}: Our analysis indicates that 5.6\% of Airflow-related questions are related to developers' limited knowledge in programming, specifically Python, and DevOps practices. Since Airflow workflows are defined through Python code, proficiency in Python is essential. Additionally, Airflow's capabilities in logging and monitoring workflows are crucial for ensuring their reliability during runtime, necessitating a comprehensive understanding of the software development lifecycle, including practices like test-driven development and DevOps methodologies. Airflow users, however, come from diverse backgrounds and vary in their experience levels, which can lead to challenges. A case in point involves a developer who faced difficulties with an Airflow task intended to consume messages from a Kafka topic via the \code{ConsumeFromTopicOperator}.\footnote{\url{https://stackoverflow.com/questions/74109768}} Despite the DAG executing without any visible errors, the anticipated messages were not reflected in the Airflow logs. This issue was ultimately attributed to a lack of understanding in Python programming
, especially in areas such as variable scopes and the application of callable functions.

\vspace{0.1cm}
\noindent \textbf{(R8) \rcseven}: The root cause of 4.8\% of the studied problems is related to Python packages or external libraries that are either not installed or improperly configured. Such issues can lead to errors and warnings that impede the system’s functionality. For instance, a developer encountered a \errorcode{No module named boto3} error in Airflow even after installing the \code{boto3} library via \code{pip}.\footnote{\url{https://stackoverflow.com/questions/61562079}} This error indicates that the boto3 library was not accessible in the Python environment where Airflow's tasks were running. To resolve such issues, the PYTHONPATH environment variable within Airflow must be correctly set to include the directories where the necessary libraries are installed. In environments with distributed architectures, where workflows extend across multiple nodes or clusters, managing environmental dependencies is especially crucial. Each node involved in the workflow must have the appropriate library versions installed to avoid conflicts and ensure a consistent and efficient workflow.


\vspace{0.1cm}
\noindent \textbf{(R9) \rcnine}: The root cause in 3.1\% of studied problems is related to the inherent complexity of the workflows being developed. These workflows often consist of multiple interconnected tasks with complex dependencies, demanding advanced and thoughtful workflow design. For instance, a developer faced challenges in accurately defining task relationships within a complex workflow that required both sequential and parallel execution patterns.\footnote{\url{https://stackoverflow.com/questions/71052585}} The specific DAG in question initially involved sequential tasks (step\_1 and step\_2), followed by the parallel execution of three tasks (X, Y, Z). The complexity was raised when the developer needed to integrate additional sequential dependencies (tasks A and B) within the parallel segment of the workflow, specifically for task X, while maintaining the parallel execution of tasks Y and Z. 

\vspace{0.1cm}
\noindent \textbf{(R10) \rcten}: 0.8\% of the Airflow-related questions we analyzed are related to breaking changes introduced in newer versions of Airflow. As Airflow continues to evolve, its APIs undergo changes, including the introduction of new ones or modifications to existing ones. Developers who are using older versions of Airflow might encounter compatibility issues if they rely on outdated APIs. For example, a developer encountered an error in Airflow 2.5.1 due to the absence of the ``execution\_date'' attribute in the TaskFail object.\footnote{\url{https://stackoverflow.com/questions/75612221}} This problem stemmed from a change made in Airflow 2.3, which deprecated the ``execution\_date'' attribute in the TaskFail class. However, the removal of this attribute wasn't clearly communicated in the error messages or the changelog for Airflow 2.3, and this breaking change took effect in version 2.5.1, causing the error.

\noindent \textbf{Unclear}: The root cause for the 19.3\% of the Airflow-related questions was labeled as unclear because these questions either lack accepted answers or
do not contain informative discussions or clarifications in their question description, answers, and comments. 


\begin{summary_RQ2}{}{}
We identify 10 types of root causes underlying the challenges that developers encounter with Airflow, with incorrect configurations in DAGs emerging as the most common issue. These root causes indicate a dual problem: firstly, there are notable gaps in Airflow's official documentation that fail to adequately guide users; secondly, there is a shortfall in developers' understanding and knowledge in terms of Airflow, external systems that interact with Airflow, general programming, and DevOps. 
\end{summary_RQ2}

\subsection{\rqthree}
\noindent\textbf{Motivation:} In RQ2, we identified a notable knowledge gap between Airflow's documentation and the needs of developers, which contributes significantly to the challenges faced in Airflow-related projects. Building on this insight, RQ3 aims to explore the common types of shared resources and websites in questions related to Airflow. Our analysis results can shed light on how external information is utilized by the community to address specific challenges and questions related to Airflow.


\subsubsection{Approach}
To address RQ3, we implemented the following methodology: 

\begin{itemize}
    \item \textbf{Step 1: Collect shared URLs.} Our first step involved extracting URLs from all components (questions, answers, comments) of all the collected 9,737 SO questions with Airflow-related tags. We utilized regular expressions to capture the URLs and then counted the frequency of URL sharing across questions, answers, and comments.
    \item \textbf{Step 2: Identify root domains.} We then automatically extracted the root domains of these URLs to understand where these resources were hosted. After analyzing these domains, we ranked them based on their frequency of occurrence to determine the most commonly referenced sources. Note that we do not consider image resources. Image hosting sites like ``i.stack.imgur.com'' were ignored. 
    \item \textbf{Step 3: Merge similar domains.} To further assess the relevance of these popular resources to Airflow, we focused on the Top-20 most commonly cited domains from Step 2, which collectively accounted for 85\% of all shared URLs. Our review revealed a variety of domains with high relevance, such as \domain{github.com} and \domain{gist.github.com}, known for hosting source code. We then consolidated similar domains to eliminate duplicates. 
Furthermore, we manually merged several domains based on their specific relevance to Airflow. For example, domains that provide documentation for external systems with which Airflow interacts were grouped into one category. Last, we name each category based on its nature and relevance to Airflow.
\end{itemize}

\subsubsection{Results}
Our analysis shows that external resources are frequently referred to in Airflow-related questions on Stack Overflow. We identified a total of \URLReferences instances of URL sharing across questions, answers, and comments. Notably, 69.53\% of all Airflow-related question posts contain at least one external URL reference. As shown in Table~\ref{tab:rq3_distribution_url}, most URLs occur within answers, highlighting that answers often rely on external resources to provide comprehensive solutions or further information. 
\vspace{0.1cm}
\begin{table*}[h]
\centering
\caption{Prevalence of the URLs in questions, answers, and comments}
\vspace{-0.2cm}
\resizebox{12cm}{!}{%
\begin{tabular}
{p{0.23\linewidth}p{0.28\linewidth}p{0.9\linewidth}}
\hline
\toprule
    \textbf{Type} & \textbf{URLs Occurrences} & \textbf{Associated Questions (\% in Airflow-related Questions)} 
    \\                                                                            
\midrule
     Comments & 3,434 & 2,268 (23.65\%)
      \\     
\midrule
     Questions & 4,533 & 2,828 (29\%)
      \\   
\midrule
     Answers & 9,092 & 4,389 (45.76\%)
      \\ 

 \bottomrule
\label{tab:rq3_distribution_url}
\end{tabular}}
\end{table*}
\vspace{-0.2cm}
      
\begin{table*}[h]
\centering
\caption{A taxonomy of top-20 most popular resources referred to in Airflow-related question posts.}
\resizebox{12cm}{!}{%
\begin{tabular}{p{0.52\linewidth}p{0.23\linewidth}p{0.29\linewidth}}
\toprule
\textbf{Category} & \textbf{\% of URLs} & \textbf{Domains} \\
\midrule
Documentation (Airflow) & 27.55\% & airflow.apache.org \\
& 1.1\% & airflow.incubator.apache.org \\
\midrule

GitHub & 24.33\% & github.com \\ 
& 0.34\% & gist.github.com \\
\midrule
Question and Answer site & 14\% & stackoverflow.com \\
\midrule

Documentation (External Systems) & 8.76\% & cloud.google.com \\
& 1.14\% & docs.aws.amazon.com \\
& 0.46\% & kubernetes.io \\
& 0.75\% & docs.docker.com  \\
& 0.32\% & aws.amazon.com \\
\midrule
Content Sharing & 1.40\% & medium.com \\
& 1.1\% & towardsdatascience.com \\
& 0.36\% & youtube.com \\

\midrule

Issue Tracking & 1.11\% & issues.apache.org \\
\midrule

Third Party Documentation & 1.82\% & docs.astronomer.io \\
& 0.34\% & www.astronomer.io \\
\midrule

Package Management & 0.81\% & pypi.org \\
\midrule

Collaboration Wiki & 0.65\% & cwiki.apache.org \\
\midrule

Documentation (Python) & 0.59\% & docs.python.org \\
\midrule

Editor for Cron Scheduling & 0.32\% & crontab.guru \\




\bottomrule
\label{tab:rq3_distribution_category}
\end{tabular}}
\end{table*}

Table~\ref{tab:rq3_distribution_category} shows the Top-20 most frequently referenced domains in Airflow-related discussions, organized into a taxonomy of 11 distinct categories. The analysis reveals that a significant 38.13\% of these domains are centered around documentation. This documentation spans a broad spectrum, encompassing Airflow itself, external systems, those reorganized by third-party, and Python-related resources.
In addition to documentation, the results indicate a high frequency of references to GitHub and various Q\&A sites.

Our analysis yields several key observations. First, Airflow's official website, \domain{airflow.apache.org}, is the most frequently cited source, accounting for 27.55\% of all URLs. This site primarily offers basic information and functionalities about Airflow. However, documentation of external systems, particularly Google Cloud services found on \domain{cloud.google.com}, is also notably prevalent, comprising 8.76\% of the references. This highlights the importance of understanding both Airflow and its integrations with external cloud services. 

The second most referenced source is \domain{github.com}. 
Our further quantitative analysis of this domain finds that among all URLs from the GitHub domain, a significant 70.3\% are associated with the Airflow repository. Within this specific group, a quarter (25\%) of the URLs point to issues or pull requests of Airflow. A further qualitative examination reveals that GitHub links are often shared to directly reference the source code and help understand how a particular feature or function is implemented that can guide to resolving an issue. Additionally, sharing links to issues and pull requests on GitHub helps identify specific problems or feature requests, facilitating collaboration on solutions and engaging the community in discussions about project direction and improvements. 

The website www.astronomer.io contributes to 1.33\% of the shared URLs. \domain{Astronomer.io} provides documentation for Astro, a cloud solution designed to enhance the management of Apache Airflow by adding extra functionalities. This suggests a trend in the community towards leveraging third-party documentation sources for extended knowledge and capabilities beyond what's available in the core Airflow documentation. 

\begin{summary_RQ3}{}{}
In Airflow-related questions on Stack Overflow, external resources are prominent, particularly in the answers. The most frequently referenced resource is Airflow's official website, indicating its central role as a primary source of information. However, our analysis also shows a significant reliance on documentation from external systems and third-party resources. This trend indicates the diverse range of information sources developers consult to address Airflow-related challenges. Moreover, the developers often turn to the official GitHub repository of Airflow for deeper and more detailed insights and solutions, highlighting its significance within the Airflow community.
\end{summary_RQ3}

\section{Discussion}\label{sec:discussion}

In this section, we begin by presenting the connection between the root causes and challenges identified in RQ1 and RQ2. Then, we discuss the actionable implications of our findings for researchers and practitioners in the field. In the end, we compare our findings with related empirical studies on developers' challenges in other domains.

\subsection{Implications}

\begin{table*}[h]
    \centering
    \caption{The Top-4 most prevalent root causes associated with the seven high-level Airflow-related challenges}.
    \resizebox{12.5cm}{!}{%
    \begin{tabular}{llcc}
        \toprule
        \textbf{Challenges} & \textbf{Root Cause} & \textbf{Percentage (\%)} \\
        \midrule
     (C1) Workflow Definition & \rcone & 36.24 \\
         & \rcthree & 13.76 \\
         & \rcfive & 10.40 \\
                  & \rcfour & 8.72 \\

                         \midrule
     (C2) Workflow Execution & \rcone & 35.33 \\
         & \rcthree & 23.95 \\
         & \rcfour & 11.98 \\
         & \rcnine & 10.18 \\
                         \midrule

         (C3) Environment Setup and Deployment & \rctwo & 44.06 \\        
                 & \rcseven & 25.87 \\
                  & \rceight & 10.49 \\
         & \rcthree & 8.39 \\

        \midrule

         (C4)  Quality Assurance and Workflow Maintenance & \rctwo & 23.85 \\
         & \rcfour & 20.18 \\
         & \rcthree & 15.60 \\
                  & \rceight & 10.39 \\

       \midrule

       (C5) Security & \rctwo & 37.50 \\
         & \rcfive & 25.00 \\
         & \rcfour & 23.53 \\
         & \rceight & 6.25 \\
        \midrule

        (C6) Optimization 
        & \rctwo & 37.50 \\ 
        & \rcthree & 25 \\
         & \rcfive & 18.75 \\
         & \rcnine & 6.25 \\
        \midrule


        (C7) Adoption of Airflow & \rcthree & 81.8 \\
         & \rcsix & 9.09 \\
         & \rcnine & 9.09 \\

       
    
        \bottomrule
    \end{tabular}}
    \label{tab:root_causes_distribution}
\end{table*}

Table~\ref{tab:root_causes_distribution} provides a breakdown of the Top-4 most prevalent root causes associated with each of the seven high-level challenge types encountered by developers while developing and managing their workflows utilizing Airflow. Our main findings are: 
\begin{itemize}
\item Three or more distinct root causes influence each of the seven challenge types, with the Top-4 root causes accounting for between 69\% and 100\% of the cases in each challenge type. This indicates a diverse range of underlying issues contributing to each challenge type.
\item In six of the seven challenge categories, a single root cause is responsible for approximately half of the questions in that category. This contrasts with challenges related to quality assurance and workflow maintenance, where the root causes are more varied and spread across multiple factors.
\item Insufficient knowledge about Airflow or external systems or Python is among the Top-4 root causes for all challenge types.
\item
Complex environmental configuration is the Top-1 root cause for four challenge types. 
\end{itemize}

Based on Table~\ref{tab:root_causes_distribution}, we make the following implications for different stakeholders. 

\subsubsection{Implications for researchers}

\noindent \textbf{Need research support on automated environment configuration.} Developers regularly encounter obstacles related to environmental configurations while working with Airflow. Table~\ref{tab:root_causes_distribution} illustrates that the intricacy of configuring environments is a principal factor that intensifies a variety of challenges, notably in "Environment Setup and Deployment" and "Security". This indicates a pressing need for future research aimed at aiding developers in the configuration of environments for workflows orchestrated by Airflow. Potential solutions could include the development of automated tools for configuring parameters or advanced debugging tools designed to streamline and simplify the configuration process. Prior studies~\citep{xu2016early} proposed approaches to tune configurations or detect configuration errors related to failure handling and fault tolerance of software systems. However, existing studies have not targeted the challenges in configuring workflows within data-intensive software systems, nor have they delved into the specifics of workflows orchestrated using Airflow.


\vspace{0.1cm}
\noindent \textbf{Need research support on operator recommendation and configuration guidance.} Developers often face significant obstacles due to incorrect workflow configurations, identified as the primary root cause for the most common challenge, i.e., ``Workflow Definition'' as shown in Table~\ref{tab:root_causes_distribution}. This issue, as explored in RQ1 and RQ2, often arises from the complexities involved in identifying appropriate operators and determining their optimal configurations. Addressing this challenge necessitates focused research on the development of sophisticated recommendation tools. Such tools would offer invaluable assistance to developers by steering them toward the selection and configuration of the most apt operators for their specific workflow needs. 

\subsubsection{Implications for Airflow application developers}
\noindent \textbf{Essential to gain a diverse skill set.} Table~\ref{tab:root_causes_distribution} reveals that a lack of comprehensive knowledge in Airflow, external systems, and Python significantly contributes to various challenges. This highlights the critical need for developers to acquire a broad spectrum of skills within the Airflow ecosystem. Developers should enhance their expertise across several domains, including Python scripting proficiency, a thorough understanding of the DAG structure, insights into distributed systems architecture, and a solid grasp of Airflow's diverse components. Adopting this knowledge enables developers to adeptly manage and leverage the full capabilities of Airflow, thereby streamlining workflow processes and overcoming the intricacies inherent in its operation.

\vspace{0.1cm}
\noindent\textbf{Proactive monitoring and engagement in the Airflow community:} The ongoing development and improvement of Airflow are propelled by its dedicated community, which boasts 2,824 contributors\footnote{\url{https://github.com/apache/airflow}}. Our findings from RQ2 highlight how breaking changes and bugs in specific releases can create obstacles for developers, especially when working with features that are not fully supported, such as dynamic workflows. To effectively overcome these challenges, we recommend that developers actively participate in the Airflow community. Engaging in forums, subscribing to mailing lists, and contributing to discussions are pivotal actions that not only ensure developers are up-to-date with the latest developments and practices but also promote a supportive and collaborative culture. This proactive involvement facilitates the sharing of valuable insights, strategies for addressing common issues, and guidance on maximizing the utility of Airflow’s features. Through such dynamic engagement, developers can both contribute to and benefit from the collective wisdom and innovative progress of the Airflow ecosystem.


\subsubsection{Implications for Airflow platform developers}

\noindent \textbf{Enable flexible operator design:} RQ1 highlights that developers frequently face challenges in task configuration (C1.1), particularly with dynamic parameterization and the development of custom operators to expand Airflow's functionality. Additionally, the proliferation of similar operators, each with its own unique configuration options, has compounded these challenges. This situation underscores the necessity for a strategic overhaul of operator design within Airflow. A potential solution is to adopt a more modular and flexible architecture. Such an architecture would not only facilitate code reuse but also ensure adherence to a consistent API framework. By implementing a design philosophy that emphasizes modularity and extensibility, Airflow can greatly enhance the developer experience, making it easier to create and manage custom operators while promoting a more streamlined and efficient workflow development process. 

\vspace{0.1cm}
\noindent \textbf{Comprehensive support for shared data storage:} Our findings in RQ1 show that incorrect workflow configuration (R1) is the primary root cause of challenges faced by developers, a significant portion of which can be traced back to Airflow's current limitations in offering sophisticated shared data storage solutions. The absence of robust shared data storage mechanisms, capable of seamless integration with external storage systems and facilitating efficient data exchange through mechanisms like XCom, compromises data consistency and reliability across tasks. To address these challenges, developers of Airflow could evolve its support for shared data storage. Enhancing its integration capabilities with external storage solutions and optimizing the efficiency of data exchange methods will be crucial steps forward. Such advancements will significantly elevate the utility and performance of Airflow in managing complex data workflows. 

\vspace{0.1cm}
\noindent \textbf{Need to enhance the official documentation of Airflow:} While Airflow's official documentation is frequently cited in responses to Airflow-related inquiries (as shown in RQ3), developers encounter difficulties in locating specific information within it.\footnote{\url{https://stackoverflow.com/questions/74640349}} To address this, efforts are needed to improve the current official website of Airflow, especially to improve the description for the configuration of the built-in and providers' operators as well as environmental configurations. The goal should be to make it not only comprehensive but also intuitively organized, ensuring that researchers and developers can quickly find the information they need to utilize Airflow's features effectively. This approach will significantly reduce the learning curve and improve the user experience.

A potential approach to improving Airflow documentation involves leveraging content from the Airflow repository. As evidenced in RQ3, components of the Airflow source code are frequently referenced, indicating their value and relevance. Large Language Models (LLMs) can be leveraged to extract and update documentation automatically based on repository changes. This approach could ensure that the documentation stays up-to-date with ongoing development, addressing an important concern among developers.

Moreover, enriching the documentation with best practices and real-world examples is vital. Such content can serve as a roadmap for developers, showing them how to leverage Airflow efficiently in their specific contexts. Practical demonstrations of Airflow in various use cases will provide valuable insights into workflow optimization, common problem resolutions, and innovative techniques for project enhancement. This strategy ensures that the documentation is not just informative but also actionable, helping users to exploit Airflow's capabilities in their operational environments fully.
\subsection{Comparing to other Domains}


Our taxonomy highlights a range of challenges exclusive to Workflows as Code, i.e., Task Definition and Configuration, Integration with External Systems, Data Sharing between Tasks, Workflow Design, Scheduling and Triggers, and Task Dependency.


There are intersections between our findings and the challenges identified in previous studies, particularly regarding security, testing, installation, performance optimization, maintenance, and logging within the context of Workflows as Code. Security concerns, as identified as one type of challenge in the configuration-as-code domain~\citep{rahman2018questions}, are crucial for safeguarding the integrity and confidentiality of workflow development. The testing challenge, discussed in the same work~\citep{rahman2018questions}, is essential for ensuring the reliability and functionality of workflow implementations. Additionally, installation challenges, highlighted in Deep Learning development~\citep{zhang2019empirical}, deployment~\citep{chen2020comprehensive}, and the configuration-as-code domain~\citep{rahman2018questions}, are crucial for effectively setting up and configuring workflow platforms. Similar to the domains of Spark~\citep{wang2022empirical} and Big Data~\citep{bagherzadeh2019going}, logging and performance optimization plays a vital role in capturing workflow-related logs and enhancing performance within the context of Workflows as Code.

It is important to note that among the studies mentioned, only Wang et al.~\citeyear{wang2022empirical} analyzed the root causes behind these challenges, like our study is doing. Two of our identified root causes closely correspond with their findings: complex environmental configuration and a steep learning curve required for the target platform.

\section{Threats to validity}\label{sec:threats}

\noindent \textbf{External Validity.} Threats to external validity are related to the
generalization of our results. Our taxonomy of challenges and root causes associated with Airflow is based on a thorough analysis of 1,000 randomly selected samples from a total of \TotalQuestionPostsScoreGteZero Stack Overflow (SO) posts tagged with Airflow-related terms. While we did not individually analyze each of the \TotalQuestionPostsScoreGteZero posts, we ensured statistical significance in our sample with a 95\% confidence level and a 3\% confidence interval. Similar to other studies that focus on understanding the challenges developers encounter within specific domains~\citep{yang2016security,rosen2016mobile, ahmed2018concurrency, alshangiti2019developing, bagherzadeh2019going,alshangiti2019developing, chen2020comprehensive}, our study exclusively relies on SO as the data source for investigating the challenges faced by developers, which may not generalized to all kinds of challenges developers face while utilizing Airflow. In the future, we plan to validate our findings on other communication forums of Airflow-related practitioners. 

Another threat to external validity arises from our choice of analysis target. As the initial effort to understand the challenges of implementing Workflows as Code, in this paper, we selected Apache Airflow from among the various platforms supporting such implementations. Our choice was primarily influenced by Airflow's widespread use in industry and its prominence on SO, where it has the highest number of question posts. In comparison, as of March 2024, the number of SO questions for other platforms is markedly lower: Luigi with 343 questions, Prefect with 192, Dagster with 171, Metaflow with 30, and Flyte with 19. Although these numbers are significantly smaller compared to Airflow, our analytical approach could be adapted for future research on these platforms, assuming more data becomes available.


\vspace{0.1cm}
\noindent \textbf{Internal Validity.} Threats to internal validity are related to experiment errors or biases. In addressing RQ1 and RQ2, we carefully annotated 1,000 SO question posts to construct two comprehensive taxonomies: one for categorizing types of challenges related to Airflow and the other for identifying their underlying root causes. To mitigate potential bias and errors introduced by the labeling process, we employed a robust two-step open coding process. Initially, two authors collaboratively labeled a representative pilot set (30\%) of the data. Following this, they independently analyzed and labeled the remaining 70\% posts. This approach allowed for the cross-validation of findings. Whenever discrepancies emerged in the labeling process, two annotators discussed and resolved them jointly. We computed the Cohen’s Kappa and found the agreement scores to be sufficiently high (0.81 and 0.79). When measuring the popularity and difficulty of the Airflow-related questions studied in the preliminary study, we opted to follow the same set of measurement metrics used in previous studies~\citep{yang2016security,rosen2016mobile, ahmed2018concurrency, alshangiti2019developing, bagherzadeh2019going}. While these metrics provide a structured approach to our analysis, it's important to acknowledge that they might not comprehensively capture all aspects of question popularity and difficulty.



\section{Conclusion}\label{sec:conclusion}

Workflows as Code simplifies the complexities of orchestrating workflows in data-intensive software systems. In this study, we gathered 9,591 Q\&A posts related to the leading Workflows as Code platform, Airflow from Stack Overflow and carried out an empirical analysis to explore the challenges and root causes encountered by developers using Airflow, as well as to examine the external resources shared through URLs within these questions. 

Manually analyzing 1,000 sampled posts, we identified 7 high-level categories and 14 sub-categories for Airflow-related challenges, with the most significant challenge lying in ``Workflow Definition'', accounting for \ChallengeWorkflowDefinition of the identified challenges. Through our analysis, we identified ten root causes underlying these challenges, ranging from incorrect workflow configuration to a lack of basic knowledge of Airflow. We also found that external resources are frequently shared in answers to Airflow-related questions, which underscores the importance of diverse knowledge required to resolve Airflow-related challenges that developers face. 

Our findings provide insights into the practical difficulties encountered when implementing Workflows as Code, underlying the necessity for improved support and a more profound comprehension of this paradigm in the software engineering field.

\bibliographystyle{elsarticle-harv} 
\clearpage

\bibliography{main}





\end{document}